# Gluon saturation and pseudo-rapidity distributions of charged hadrons at RHIC energy regions*

WEI Xinbing[1], FENG Shengqin [1, 2, 3+]

[1] College of Science, China Three Gorges University, Yichang 443002, China

[2] Key Laboratory of Quark and Lepton Physics (Huazhong Normal Univ.), Ministry of Education, Wuhan 430079, China

[3] School of Physics and Technology, Wuhan University, Wuhan 430072, China

We modified the gluon saturation model by rescaling the momentum fraction according to saturation momentum and introduced the Cooper-Frye hydrodynamic evolution to systematically study the pseudo-rapidity distributions of final charged hadrons at different energies and different centralities for Au-Au collisions in relativistic heavy-ion collisions at BNL Relativistic Heavy Ion Collider (RHIC). The features of both gluon saturation and hydrodynamic evolution at different energies and different centralities for Au-Au collisions are investigated in this paper.

## 1. Introduction

After several years of RHIC operation, lots of experimental results on multi-particle productions become available. It appears that the experimental data on hadron multiplicity and its energy, centrality, and rapidity dependence so far are consistent with the result [1,2] based on the ideas of gluon saturation [3,4] or the color glass condensate (CGC) [5-10].

CGC is a state that at very high energies, a new form matter, a dense

* Supported by the National Natural Science Foundation of China (No. 10975091) Excellent Youth Foundation of Hubei Scientific Committee (2006ABB036) and Education Commission of Hubei Province of China (Z20081302)

+Corresponding author: fengsq@ctgu.edu.cn

condensate of gluons is created. In a hadron, the constituents are the gluons, valence quarks and sea quarks. As the collision energy increases, the gluons in a hadron will radiate new gluons and the gluons take a dominate state.  But the number of a certain kind of gluons will not increase for ever in a fixed hadron, and then it approaches toward a constant. That is gluon saturation.

There exist a lot of extensive work[1-10] for the description of gluon productions in nuclear collisions in the saturation regime.   One can find that the nonlinear effects of saturation region become important. Perturbative solutions for the collision of two nuclei of the MV model were obtained in Refs. [7-17].  One of our purposes in this paper is to investigate the rapidity dependence of final hadrons.

Here we should mention a novel gluon saturation model proposed by Kharzeev, Levin, and Nardi (KLN) [1,2,18,19] to discuss the gluon saturation mechanism and calculate the gluon rapidity distribution.  An analytical scaling function which embodies the predictions of high density QCD on the energy, centrality and rapidity dependences of hadron multiplicities in nuclear collisions are proposed in this model[1,2,18,19].  In Ref.1, it is found that the simplified KLN model could fit well the central rapidity distribution ($-4.5 < \eta < 4.5$) as shown in Fig.1(a) when compared with the RHIC data[20].  But if we extend the simplified KLN model to fit the distribution of full rapidity region as shown in Fig.1(b), it is found that the simplified KLN model jumps abruptly at large rapidity region.  In order to solve the problem, we modified the gluon saturation model by rescaling the momentum fraction according to saturation momentum $Q_s$.  And then we introduce the Cooper-Frye hydrodynamic evolution to study the rapidity distributions of final charged hadrons in this paper.

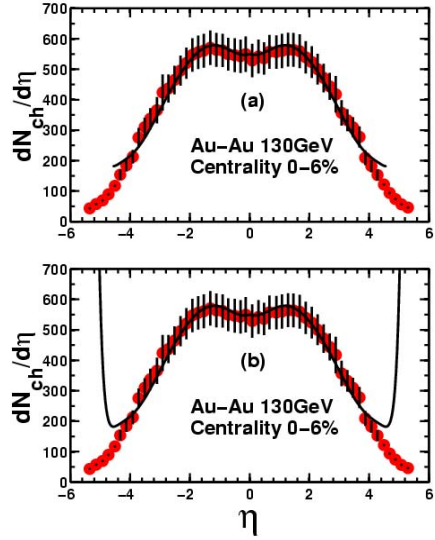

Fig.1 The charged hadron pseudo-rapidity distribution of central collision at $\sqrt{s_{NN}}$ = 130 GeV. Fig.1(a) comes from the simplified KLN model[1] and Fig.1(b) extends the simplified KLN model[1] to large rapidity region. The real lines are the calculated results and the experimental results are from data of Phobos/RHIC[20]

This paper is organized as follows. In Sec.2, we give a detailed review of the modified gluon saturation model for the initial condition, and then introduce the Cooper-Frye hydrodynamic evolution. The comparison and analysis of charged hadron pseudo-rapidity distribution of RHIC with the results of the model are given in Sec.3 . Sec.4 gives a summary and conclusions.

## 2. The modified gluon saturation model connecting with hydrodynamics

As mentioned above, color glass condensate creates a dense condensate of gluons[21]. There is a very high density of mass-less gluons and these gluons can be packed until their phase space density is so high that interactions prevent more gluon occupation. The gluons occupy higher momentum with increasing energy, so that the coupling becomes weak. The gluon density saturates at a value of order $\alpha_s \ll 1$, corresponding to a multi-particles state which is a Bose condensate.

One can define the transverse momentum of gluon saturation as the saturation momentum $Q_s$ (saturation scale) at very high collision energy. The saturation scale can be estimated as[11-17]:

$$Q_s^2 = \alpha_s N_c \frac{1}{\pi R^2} \frac{dN}{dy} \qquad (1)$$

KLN[1,2] discussed gluon saturation of two identical nuclei collision situation and introduced two auxiliary variables

$$Q_{s,\min}(y, \sqrt{s_{NN}}) = \min\{Q_s(A; x_1), Q_s(A; x_2)\} \qquad (2)$$

$$Q_{s,\max}(y, \sqrt{s_{NN}}) = \max\{Q_s(A; x_1), Q_s(A; x_2)\} \qquad (3)$$

From Eq.2 and Eq.3, when y=0, we can find $Q_{s,\max} = Q_{s,\min}$, and when $y>0$ (or $y<0$), $Q_{s,\min} = Q_s(A, \mp y, \sqrt{s_{NN}})$, $Q_{s,\max} = Q_s(A, \pm y, \sqrt{s_{NN}})$. Then one can define[1,2]

$$Q_{s,\min} = Q_s(A, -y, \sqrt{s_{NN}}) = Q_s(\sqrt{s_{NN0}})(\sqrt{s_{NN}}/\sqrt{s_{NN0}})^{\lambda/2} e^{-|y|/2} = Q_s(s_{NN})e^{-|y|/2} \qquad (4)$$

$$Q_{s,\max} = Q_s(A, y, \sqrt{s_{NN}}) = Q_s(\sqrt{s_{NN0}})(\sqrt{s_{NN}}/\sqrt{s_{NN0}})^{\lambda/2} e^{|y|/2} = Q_s(s_{NN})e^{|y|/2} \qquad (5)$$

where $\sqrt{s_{NN0}} = 130$GeV, $Q_s^2(\sqrt{s_{NN0}}) = 2.05$ [1,2], and $\lambda$ the growth of the gluon structure functions at small x in deep-inelastic scattering. The HERA data are fitted with $\lambda \approx 0.25 - 0.3$ [22-24].

The differential cross section of gluon production in A-A collision can be written down as [3-4]:

$$E\frac{d\sigma}{d^3 p} = \frac{4\pi N_c}{N_c^2 - 1} \frac{1}{p_t^2} \int dk_t^2 \alpha_s \left[ \varphi_{A_1}(x_1, k_t^2) \times \varphi_{A_2}(x_2, (p-k)_t^2) \right] \qquad (6)$$

where $k_t$ and $p_t$ are the transverse momentums of parton in a hadron before collision and of the produced gluon, respectively, and $\alpha_s$ is the running coupling coefficient and takes a smaller value. Unintegrated gluon distribution $\varphi(x, k_t^2)$ describes the probability to find a gluon with a given $x$ and transverse momentum $k_t$ inside the nucleus A[1,2,18,19]. The gluon saturation function can be given by

$$xG(x, p_t^2) = \int^{p_t^2} dk_t^2 \varphi(x, k_t^2) \ . \qquad (7)$$

The expression shows that gluon saturation function is the average value of

$\varphi(x, k_t^2)$ at a given transverse momentum region. Therefore, the multiplicity distribution of gluons is[1,2]

$$\frac{dN_g}{dy} = \frac{1}{S}\int d^2 p_t \left( E\frac{d\sigma}{d^3 p} \right) \qquad (8)$$

S is either the inelastic cross section for the minimum bias multiplicity, or a fraction of it corresponding to a specific centrality cut[1,2].

KLN[1] gave the gluon distribution by considering two integration regions: $k_t \ll p_t$ and $|\vec{p}_t - \vec{k}_t| \ll p_t$; this leads to

$$\frac{dN_g}{dy} = \frac{1}{S}\int d^2 p_t \left( E\frac{d\sigma}{d^3 p} \right) = \frac{1}{S}\frac{4\pi N_c \alpha_s}{N_c^2 - 1}\int \frac{d^2 p_t}{p_t^2}\left[ \varphi_{A_1}(x_1, p_t^2)\int^{p_t} dk_t^2 \varphi_{A_2}(x_2, k_t^2) \right.$$
$$\left. + \varphi_{A_2}(x_2, p_t^2)\int^{p_t} dk_t^2 \varphi_{A_1}(x_1, k_t^2) \right] \qquad (9)$$
$$= \frac{1}{S}\frac{4\pi N_c \alpha_s}{N_c^2 - 1}\int_0^\infty \frac{d^2 p_t}{p_t^4} x_2 G_{A_2}(x_2, p_t^2) x_1 G_{A_1}(x_1, p_t^2) \quad .$$

The gluon distribution takes as[1,2,18,19],

$$xG(x, p_t^2) = \begin{cases} \dfrac{\kappa}{\alpha_s(Q_s^2)} S p_t^2 (1-x)^4 & p_t < Q_s(x), \\ \dfrac{\kappa}{\alpha_s(Q_s^2)} S Q_s^2 (1-x)^4 & p_t > Q_s(x), \end{cases} \qquad (10)$$

According to Ref.[18,19], the $p_t$ integration is divided into three regions:

(1). $p_t < Q_{s,min} < Q_{s,max}$, in this region, both parton densities for the two nuclei are in the saturation region.

(2). $Q_{s,min} < p_t < Q_{s,max}$, in this region, one nucleus is in saturation region, and the other one is in the normal DGLAP region.

(3). $p_t > Q_{s,max} > Q_{s,min}$, in this region the parton densities in both nuclei are in the DGLAP evolution region.

The gluon distribution is given as follows[1,2]:

$$\frac{dN_g}{dy} = \frac{N_c \kappa^2 \beta_0}{N_c^2 - 1} \frac{N_{part}}{Q_{s0}^2} \ln \frac{Q_{s,min}^2}{\Lambda_{QCD}^2} \left\{ \int_0^{Q_{s,min}} (1-x_1)^4 (1-x_2)^4 d^2 p_t + \right.$$

$$\left. Q_{s,min}^2 \int_{Q_{s,min}}^{Q_{s,max}} \frac{1}{p_t^2} (1-x_1)^4 (1-x_2)^4 d^2 p_t + Q_{s,min}^2 Q_{s,max}^2 \int_{Q_{s,max}}^{\infty} \frac{1}{p_t^4} (1-x_1)^4 (1-x_2)^4 d^2 p_t \right\}$$

(11)

where $\beta_0 = 11 - \frac{2}{3} N_f$, $N_f = 3$.

The KLN model[1] realized that the gluon distributions are the final hadron distributions. This was based on the assumption that the final state interactions did not change significantly the multiplicities of partons resulting from the early stages of the process[1,2].

In the KLN model[1], they defined the momentum fraction as follows:

$$x_1 = \frac{p_t}{\sqrt{s_{NN}}} e^{-y}, \quad x_2 = \frac{p_t}{\sqrt{s_{NN}}} e^{y} \quad (12)$$

Here we should say a few words about Fig.2. Even if we use the integrated Eq. 9 and Eq.11 to fit the PHOBOS experimental data of RHIC energy region, we find that the integrated KLN model cannot fit well with the experimental results at large pseudo-rapidity region as shown in Fig. 2. In order to solve the problem, it is assumed that the momentum fraction at Eq.9 should depend on the structure function and saturation momentum of the collision nuclei in the saturation region.

Here we rescale the momentum fraction according to the saturation momentum:

$$x_1 = \frac{Q_{s,min}}{\sqrt{s_{NN}}} = \frac{p_t}{\sqrt{s_{NN}}} e^{-|y|/2}, \quad x_2 = \frac{Q_{s,max}}{\sqrt{s_{NN}}} = \frac{p_t}{\sqrt{s_{NN}}} e^{|y|/2} \quad (13)$$

After carefully studying the KLN model and the recent PHOBOS experimental data, we propose to rescale the fraction momentum according to the saturation momentum, and introduce the CGC as initial condition to connect with Cooper-Frye hydrodynamic evolution to study the pseudo-rapidity distributions of final charged hadrons.

At RHIC, two collider nuclei will be Lorentz contract. At the moment after collision, they will undergo a quantum fluctuation ($\tau \sim 0-0.1\,fm/c$), density fluctuation and thermalization($\tau \sim 0.1-1\,fm/c$). For central collision of two big nuclei, we assume that fluid near the collision axis moves longitudinally and homogeneously. One can take the two nuclei as two thin pancakes, the fluid midway between the receding pancakes remains at rest[25,26].

The system of gluons initially produced from the CGC reaches a kinematically as well as chemically local equilibrate state at a short time scale. One can assume that during thermalization, the shape of the rapidity distribution is not changed. Thus, we take the initial conditions from gluon distribution obtained in the previous subsection based on the CGC.

We adopt Bjorken`s assumption[25] that $y = \eta_s$, at which y is the energy-momentum rapidity, and $\eta_s$ is the space-time rapidity ($\eta_s = \frac{1}{2} ln \frac{x^+}{x^-}$). Due to Formula (13), the gluon density[25] at point $(\tau_0, y)$ is

$$n_g(\tau_0, y) = \frac{dN_g}{\tau_0 dy} \qquad (14)$$

where $\tau_0$ is the initial time. The relations among thermodynamic variables, i.e., temperature $T$ and related number density $n(\tau_0, y)$ for the massless free parton system are as follows:

$$n(\tau_0, y) = (3/4 g_q + g_g) \frac{\zeta(3)}{\pi^2} T^3(\tau_0, y), \qquad (15)$$

then

$$T(\tau_0, y) = \left[ \frac{\pi^2 n(\tau_0, y)}{43\zeta(3)} \right]^{1/3} \qquad (16)$$

where $g_q = 2N_c N_s N_f = 36$, $g_g = 2(N_c^2 - 1) = 16$, $\zeta(3) = 1.20206$, color number $N_c = 3$, flavor number $N_f = 3$ and spin number $N_s = 2$. Cooper-Frye assumed that the particle distributions could be described by either a Bose or a Fermi distribution according to the type of the observed particle[27]. The invariant single-particle distribution of gluon in momentum space is

$$E\frac{d\sigma}{d^3p} = \int_\sigma f(x,p) p^\mu d\sigma_\mu = \frac{g}{(2\pi)^3} \int \frac{p^\mu d\sigma_\mu}{e^{E/T}-1} \quad . \tag{17}$$

Adopting a cylinder with radius R and length $2\eta_0$ in the case of no transverse fluid, one can get

$$d\sigma_\mu = \int d\eta_s \pi R^2 \tau (\cosh\eta_s, 0_\perp, \sinh\eta_s). \tag{18}$$

In the midway $\eta_s \approx 0$, therefore

$$p^\mu d\sigma_\mu = p^0 \pi R^2 \tau d\eta_s \quad , \tag{19}$$

The rapidity distribution is:

$$\frac{dN_{ch}}{dy} = \frac{g}{(2\pi)^3} \int E\frac{dN^3}{d^3p} d^2 p_t = \frac{g}{(2\pi)^3} \int_{-\eta_0}^{\eta_0} d\eta_s \cdot \int_0^\infty dp_t p_t^2 \pi R^2 \tau \cosh(\eta_s - y) f(\vec{p}) \tag{20}$$

where $f(\vec{p}) = \dfrac{1}{\exp[p_t \cosh(\eta_s - y)/T]-1}$. Consequently,

$$\frac{dN_{ch}}{dy} = \frac{g}{(2\pi)^3} 4\pi^2 R^2 \tau T^3 \left[\tanh(y+\eta_0) - \tanh(y-\eta_0)\right] \tag{21}$$

Substituting Formulas (14) and (16) into (21), one can take the following formula of rapidity distribution of final charged hadrons as follows:

$$\frac{dN_{ch}}{dy} = \frac{\pi R^2 g}{86\zeta(3)} \cdot \frac{\tau}{\tau_0} \cdot \frac{dN_g}{dy} \times \left[\tanh(y+\eta_0) - \tanh(y-\eta_0)\right] \tag{22}$$

where degeneracy $g = \dfrac{3}{4} g_q + g_g$, $\eta_0$ is the kinematical range of collective flow in the longitudinal direction of nuclear-nuclear collisions[28-34], R is the radius of the phase space of cylinder and $\tau$ is the "proper time" defined as[25]:

$$\tau = \sqrt{t^2 - z^2} \tag{23}$$

From (11) and (22), we can take the rapidity distribution of charged hadrons as follows:

$$\frac{dN_{ch}}{dy} = \frac{\pi R^2 g}{86\zeta(3)} \cdot \frac{\tau}{\tau_0} \frac{N_c \kappa^2 \beta_0}{N_c^2 - 1} \frac{N_{part}}{Q_{s0}^2} \cdot ln\frac{Q_{s,min}^2}{\Lambda_{QCD}^2} \left\{ \int_0^{Q_{s,min}} (1-x_1)^4 (1-x_2)^4 d^2 p_t + \right.$$

$$Q_{s,min}^2 \int_{Q_{s,min}}^{Q_{s,max}} \frac{1}{p_t^2}(1-x_1)^4 (1-x_2)^4 d^2 p_t + Q_{s,min}^2 Q_{s,max}^2 \int_{Q_{s,max}}^{\infty} \frac{1}{p_t^4}(1-x_1)^4 (1-x_2)^4 d^2 p_t \right\}$$

$$\times [tanh(y+\eta_0) - tanh(y-\eta_0)]$$

(24)

The pseudo-rapidity distribution takes as follows:

$$\frac{dN}{d\eta} = \sqrt{1 - \frac{m^2}{m_t^2 \cosh^2 y}} \frac{dN}{dy}, \qquad (25)$$

here $y = 0.5 \ln \frac{\sqrt{p_t^2 \cosh^2 \eta + m^2} + p_t \sinh \eta}{\sqrt{p_t^2 \cosh^2 \eta + m^2} - p_t \sinh \eta}$, $m_t^2 = p_t^2 + m^2$, $p_t = Q_s$.

### 3. The pseudo-rapidity distributions of charged hadrons

The study of relativistic heavy-ion collisions is the only known method of creating and studying, in the laboratory, systems with hadronic or partonic degrees of freedom at extreme energy and matter density over a significant volume. It is for this reason that, in recent years such studies have attracted much experimental and theoretical interest, in particular with the likelihood that, at higher energies, a new state of QCD matter is created.

The PHOBOS Collaboration, working at RHIC, has provided considerable experimental data [35] of different energies and different centralities of Au - Au collisions at $\sqrt{s_{NN}}$ =19.6, 62.4,130 and 200 GeV, respectively. This extensive body of experimental data on the global properties of particle production in heavy-ion collisions can be utilized to present insight into both our understanding of the mechanisms of particle production and the properties of matter that exist at extremes of energy and matter densities. In this paper, we will utilize the modified gluon saturation model to study the rapidity distributions of final charged hadrons.

The results are presented in Fig. 2. One finds that the calculation results from our model are consistent with those of the experimental data, especially at relatively large collision energies of $\sqrt{s_{NN}}$ =130 and 200 GeV. It is suggested that the modified gluon saturation model prefers the relatively large collision energies of $\sqrt{s_{NN}}$ =130 and 200 GeV to the relatively low collision energies of $\sqrt{s_{NN}}$ =19.6, 62.4 GeV. It is suggested that large collision energy of RHIC could easily reach the situation of gluon saturation and hydrodynamic evolution.

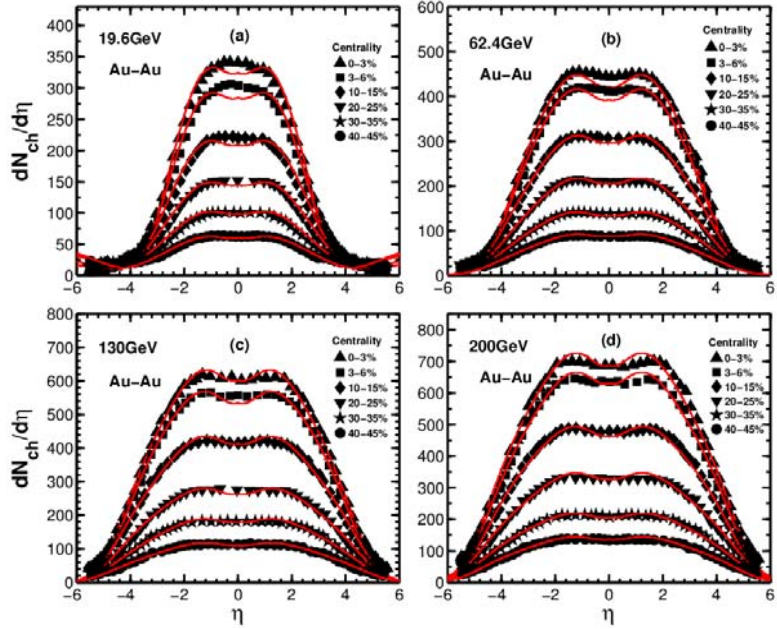

FIG. 2. The charged hadron pseudo-rapidity distributions are shown in Fig.2(a,b,c and d) for Au-Au collisions at different collision energies $\sqrt{s_{NN}}$ =19.6, 62.4,130 and 200 GeV , respectively. The solid lines are the results from our modified gluon saturation model. The experimental data are given by PHOBOS [35].

The charged hadron pseudo-rapidity distributions are shown in Fig.3 (a, b, c and d) for Au - Au collisions at different energies and different centralities of 0%-3%, 3%-6%, 35%-40% and 40%-45%, respectively. It is shown that the modified gluon saturation model tends to explain more likely of the central collision than that of the peripheral collision. The experimental data are given by PHOBOS [35]. It is suggested that the central collision system could easily reach the situation of gluon saturation and hydrodynamic evolution.

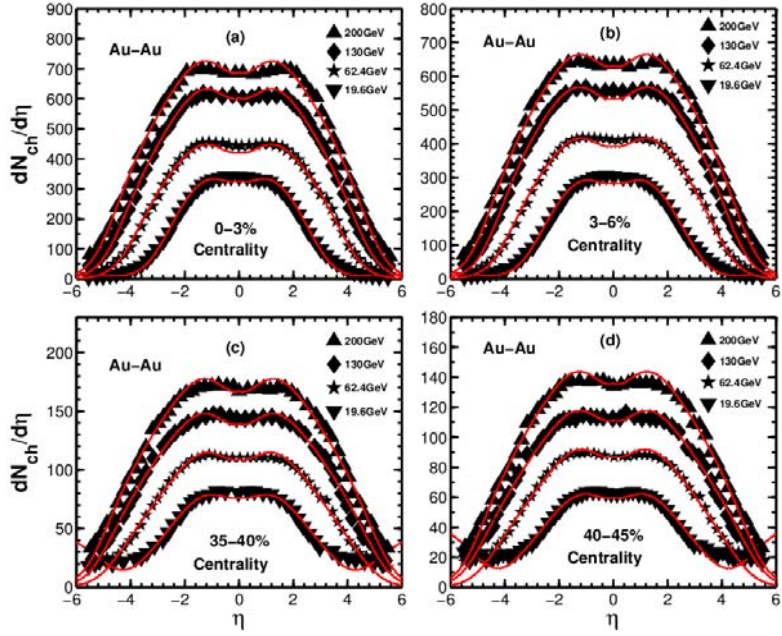

FIG. 3. The charged hadron pseudo-rapidity distributions are shown in Fig.3(a,b,c and d) for Au－Au collisions at different energies and different centralities of 0%-3%,3%-6%,35%-40% and 40%-45% , respectively. The solid lines are the results from our modified gluon saturation model. The experimental data are given by PHOBOS [35].

Here, we should say a few words about $\eta_0$, according to Bjorken evolution picture $\eta_0$ is the boost invariance limitation of thermalization source. Combining with this paper, we realize that $\eta_0$ is the rapidity limitation of emission source, where the final charged hadrons hydrodynamic evolution is from. It seems reasonable to take the limitation $\eta_0$ of emission source as the thermalization limitation of abundance of the saturation gluons.

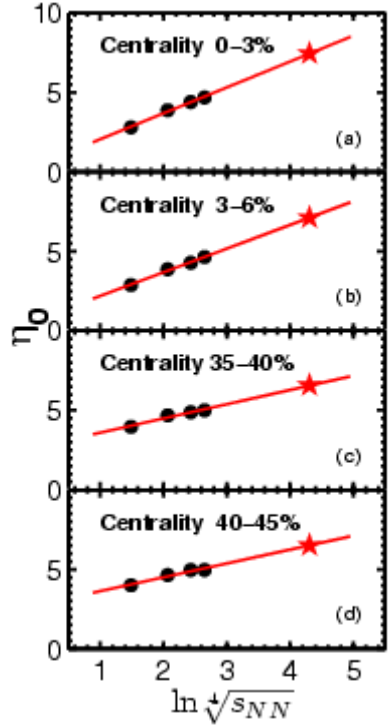

FIG. 4. The dependence of the limitation $\eta_0$ of thermalization region with $\ln\sqrt[4]{s_{NN}}$ at different centralities of 0%-3%, 3%-6%, 35%-40% and 40%-45%. The star (★) is the results of our model's prediction of LHC $\sqrt{s_{NN}}$ =5500GeV.

By fitting with the PHOBOS data[34], we get the dependence of the limitations $\eta_0$ on $\ln\sqrt[4]{s_{NN}}$ at different centrality of 0%-3%, 3%-6%, 35%-40% and 40%-45%, respectively at Fig. 4(a, b, c and d). The different linear dependences of centralities of 0%-3%, 3%-6%, 35%-40% and 40%-45%, respectively, on CMS collision energies can be given as follows:

$$\eta_0 = 1.63\ln\sqrt[4]{s_{NN}} + 0.42 \quad \text{(centrality 0%-3%)} \tag{26}$$

$$\eta_0 = 1.49\ln\sqrt[4]{s_{NN}} + 0.70 \quad \text{(centrality 3%-6%)} \tag{27}$$

$$\eta_0 = 0.90\ln\sqrt[4]{s_{NN}} + 2.70 \quad \text{(centrality 35%-40%)} \tag{28}$$

$$\eta_0 = 0.87\ln\sqrt[4]{s_{NN}} + 2.76 \quad \text{(centrality 40%-45%)} \tag{29}$$

From the linear $\ln \sqrt[4]{s_{NN}}$ relationship, we can predict the $\eta_0$ at LHC $\sqrt{s_{NN}} = $ 5500 GeV at a certain centralities. Until now, we have not found the LHC data of rapidity distributions of produced hadrons in the central region or full rapidity region of Pb-Pb collisions. We will discuss the rapidity distribution at LHC energy region in next work.

## 4. Summary and conclusion

We modified the gluon saturation model by rescaling the momentum fraction according to the saturation momentum, and introduced the Cooper-Frye hydrodynamic evolution to study the pseudo-rapidity distributions of final charged hadrons in this paper. It is found that our new modified gluon saturation model can fit well the full rapidity region of the recent published PHOBOS results at RHIC at different centralities and different energies.

From the discussions, we can find that our calculation results are consistent with those of the experimental data, especially at relatively large collision energies of $\sqrt{s_{NN}}=130$ and 200 GeV. It is suggested that large collision energy of RHIC can easily reach the situation of gluon saturation and hydrodynamic evolution.

By comparing with experimental data, we also find that the gluon saturation model prefers the central collisions to peripheral collisions. It is suggested that central collisions at RHIC can easily reach the situation of gluon saturation and hydrodynamic evolution than that of peripheral collisions. By connecting with hydrodynamic evolution, we can find that the limitations $\eta_0$ of thermalization source increase linearly with $\ln \sqrt[4]{s_{NN}}$ at different centralities.